\documentclass[%
reprint,
amsmath,amssymb,
aps,
prb,
]{revtex4-2}
\usepackage{natbib}
\usepackage{graphicx}
\usepackage{dcolumn}
\usepackage{bm}
\usepackage{color}
\usepackage{subfigure}
\begin{document}
	\preprint{APS/123-QED}
	
	\title{Impact of Charge Density Waves on Superconductivity and Topological Properties in AV$_3$Sb$_5$ Kagome Superconductors}
	\author{Xin Lin$^1$}
	\author{Junkang Huang$^1$}
	\author{Tao Zhou$^{1,2}$}
	\email{Corresponding author: tzhou@scnu.edu.cn}	
\affiliation {$^1$Guangdong Basic Research Center of Excellence for Structure and Fundamental Interactions of Matter, Guangdong Provincial Key Laboratory of Quantum Engineering and Quantum Materials, School of Physics, South China Normal University, Guangzhou 510006, China\\
$^2$Guangdong-Hong Kong Joint Laboratory of Quantum Matter, Frontier Research Institute for Physics, South China Normal University, Guangzhou 510006, China}
	\begin{abstract}

We investigates the electronic structure and superconducting gaps in the charge density wave (CDW) states of vanadium-based Kagome superconductors AV$_3$Sb$_5$, focusing on the concurrent presence of CDW and superconducting orders. Two predominant CDW configurations are explored: the trihexagonal (TrH) and star-of-David (SoD) patterns, involving charge bond order (CBO) and chiral flux phase (CFP), corresponding to real and imaginary bond orders. In the isotropic $s$-wave superconducting state, the presence of CBO alone maintains an isotropic superconducting gap, whereas the introduction of CFP induces anisotropy in the gap, manifesting time-reversal symmetry breaking due to the CFP. Our analysis extends to the topological properties of these states, revealing a marked topological phase transition in the TrH configuration from a trivial to a non-trivial state with increasing CFP intensity. This transition suggests that the introduction of CFP could catalyze the emergence of topological superconductivity, potentially leading to the presence of Majorana excitations. The results contribute significantly to understanding the complex interplay between various CDW patterns and superconductivity in Kagome superconductors. They provide a theoretical framework for the diverse experimental observations of energy gaps and open new avenues for research into topological superconductivity and its potential applications. This study underscores the necessity for further experimental and theoretical exploration to unveil novel interwinded quantum states and functionalities in these intriguing materials.

	\end{abstract}
	
	\maketitle

	\section{Introduction}

	The kagome lattice offers a compelling lattice structure for delving into unconventional states. Specifically, the family of vanadium-based kagome superconductors, AV$_3$Sb$_5$ (where A = K, Rb, or Cs), has garnered significant research interest as a robust platform to investigate the interplay between topology, charge density waves (CDW), and superconductivity~\cite{PhysRevLett.125.247002,PhysRevB.109.104512,r3,Natl.Sci.Rev.10.199,PhysRevMaterials.5.034801,r5}.
	
	Current research efforts are extensively focused on elucidating the complex characteristics of the CDW state in these materials. Experimental confirmation of CDW order has been achieved through various techniques including transport anomalies~\cite{PhysRevMaterials.5.034801,r5,PhysRevLett.127.207002,Tungsten.5.300,Nature.611.461,PhysRevB.105.205104,PhysRevB.105.L201109}, anomalous Hall effects~\cite{sciadv.abb6003.10.1126,PhysRevLett.127.236401,PhysRevB.104.L041103}, and scanning tunneling microscopy (STM)~\cite{Nature.599.216,Chen2021,PhysRevX.11.031026}, with the CDW transition temperatures ($T_{\mathrm{CDW}}$) observed to range between 78 and 103 K. Notably, the detection of quantum anomalous Hall effect suggests a breaking of time-reversal symmetry, a phenomenon further corroborated by additional experimental approaches such as muon spin spectroscopy~\cite{Nature.602.245,Nat.Commun.14.153,PhysRevResearch.4.023244}, optical polarization rotation~\cite{PhysRevB.106.205109,Nat.Commun.14.5326}, chiral transport measures~\cite{Nature.611.461,PhysRevB.105.205104}, magneto-optical Kerr effect~\cite{Nat.Phys.18.1470,PhysRevB.109.014433}, and extensive STM analysis~\cite{PhysRevB.104.075148,PhysRevB.104.035131}. From a theoretical perspective, the CDW state is believed to stem from bond-centered ordering, including both real and imaginary components~\cite{PhysRevB.104.165136,PhysRevB.104.045122,Wilson2024,fu2024exotic}. The real component involves the charge bond order (CBO)~\cite{PhysRevLett.110.126405,PhysRevB.87.115135,PhysRevB.104.035142}, while the imaginary component, the chiral flux phase (CFP)~\cite{r4,PhysRevLett.131.086601,PhysRevB.108.184506,PhysRevB.107.064506}, is primarily responsible for the observed breaking of time-reversal symmetry.
	
	The CDW exhibits a three-dimensional translational symmetry breaking characterized by a $2\times 2\times 2$ or $2\times 2\times 4$ modulation~\cite{PhysRevX.11.031026,PhysRevX.11.041030,PhysRevB.105.195136,PhysRevMaterials.7.024806} . Within the Kagome layer, the $2\times 2$ in-plane component is pivotal for understanding the physical properties of these systems. It manifests as a three wave-vector (3Q) breathing mode with patterns like the "Star-of-David" (SoD) or "Tri-Hexagonal" (TrH)~\cite{PhysRevLett.127.046401,PhysRevB.104.214513,Han2022,Subires2023,uykur2022optical,PhysRevB.107.184106,PhysRevResearch.5.L012017,Kang2023ER,PhysRevMaterials.7.024806,luo2022possible,Feng2023}. First-principles calculations indicate that the TrH state is energetically more favorable~\cite{PhysRevLett.127.046401}, a finding supported by subsequent experimental studies~\cite{PhysRevResearch.5.L012017,Kang2023ER}. Additionally, the SoD state, while being a local energy minimum, remains stable and detectable in certain vanadium-based kagome materials~\cite{PhysRevMaterials.7.024806,luo2022possible,Feng2023}.
	
	Understanding the pairing symmetry is crucial for elucidating the superconductivity mechanism in AV$_3$Sb$_5$ superconductors. Various theories have proposed different pairing symmetries~\cite{r31,r30,r28,r29,PhysRevB.106.014501,PhysRevB.109.054504}. Experimentally, different results were also reported. Measurements of thermal conductivity in a zero magnetic field have revealed a nodal superconducting gap~\cite{zhao2021nodal}. Also, STM experiments suggest the presence of residual zero-energy states and V-shaped gaps, indicative of nodal points~\cite{Chen2021}. Conversely, studies on spin-lattice relaxation rates~\cite{Duan2021}, magnetic penetration depth~\cite{Mu_2021}, impurity effects~\cite{ZhangWei2023,Roppongi2023,PhysRevLett.127.187004}, and transport measurements~\cite{wu2022nonreciprocal} support a fully gapped superconductor. Recent angle-resolved photoemission spectroscopy (ARPES) experiments have detected a nearly isotropic superconducting gap across the Fermi surface, reinforcing the case for a fully gapped kagome superconductor~\cite{Zhong2023}. The observed energy gap ratio $2\Delta/k_B T_c \approx 3.44$ aligns closely with conventional, phonon-mediated superconductors. However, there have also been reports suggesting the superconducting gap is anisotropic~\cite{Mu_2021,wu2022nonreciprocal,PhysRevLett.127.187004,gupta2022microscopic}. These contradictory experimental outcomes highlight the need for further theoretical investigation to fully comprehend the diverse characteristics of the superconducting gap in AV$_3$Sb$_5$ superconductors, including its nodal, isotropic fully gapped, and anisotropic properties.
	
	Despite potential phonon-mediated superconductivity, the presence of CDW orders suggests distinct physical properties in the superconducting state of AV$_3$Sb$_5$ materials. The coexistence of the CDW and superconducting states particularly influences the superconducting gap~\cite{hu2024discovery,mine2024direct,PhysRevLett.132.186001}, necessitating systematic studies to clarify their interplay. In this paper, we theoretically investigate the impact of in-plane CDW order on superconducting properties within a two-dimensional kagome lattice system. By considering both the TrH and SoD CDW patterns, as well as CBO and CFP orders, we analyze the effective energy gaps, spectral function, and density of states (DOS) to ascertain the influence of CDW on superconductivity. Our results demonstrate that CDW parameters markedly shape the outcome, providing theoretical insight into the diverse experimental observations. Additionally, we explore potential topological features, suggesting that the CFP within the TrH pattern could induce topological states in the superconducting phase.
	
	The remainder of this paper is structured as follows: Section II outlines the model and introduces the relevant formalism. Section III details our numerical calculations and discusses the results obtained. Finally, a brief summary is provided in Section IV.

\section{Model And Formalism}

We investigate a $2 \times 2$ CDW on a two-dimensional kagome lattice. In the CDW state, the Brillouin zone is reduced to one quarter of its original size, and each unit cell contains $12$ lattice sites. To explore the fundamental physical effects of the CDW and its impact on superconductivity, we employ an effective single-orbital model. The model comprises the normal state term $H_N$, the superconducting pairing term $H_{SC}$, and the CDW state term $H_{CDW}$, expressed as,
\begin{equation}
	\hat{H}=\hat{H}_{N}+\hat{H}_{CDW}+\hat{H}_{SC}.
	\label{eq1}
\end{equation}

The normal state Hamiltonian $H_N$, includes the nearest-neighbor (NN) hopping term and the chemical potential term, presented as,

\begin{equation}
	\hat{H}_N=-t\sum_{\langle\textbf{i}\textbf{j}\rangle \sigma}\hat{c}_{\textbf{i}\sigma}^{\dagger}\hat{c}_{\textbf{j}\sigma}-\mu\sum_{\textbf{i} \sigma}\hat{c}_{\textbf{i}\sigma}^{\dagger}\hat{c}_{\textbf{i}\sigma},
	\label{eq2}
\end{equation}
where $t$ and $\mu$ denote the NN hopping constant and the chemical potential, respectively.

\begin{figure}
	\centering
	\subfigure{\includegraphics[width=0.23\textwidth]{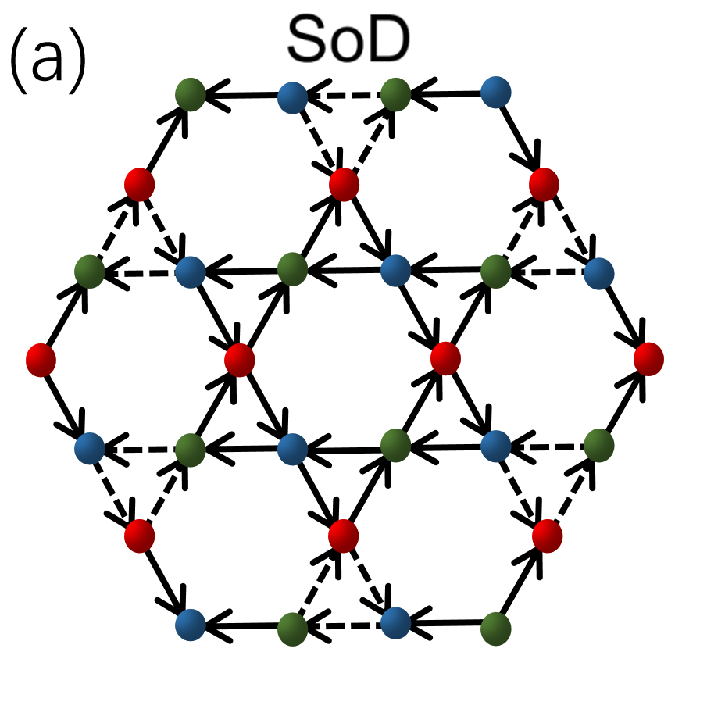} }\hspace{0mm}
	\subfigure{\includegraphics[width=0.23\textwidth]{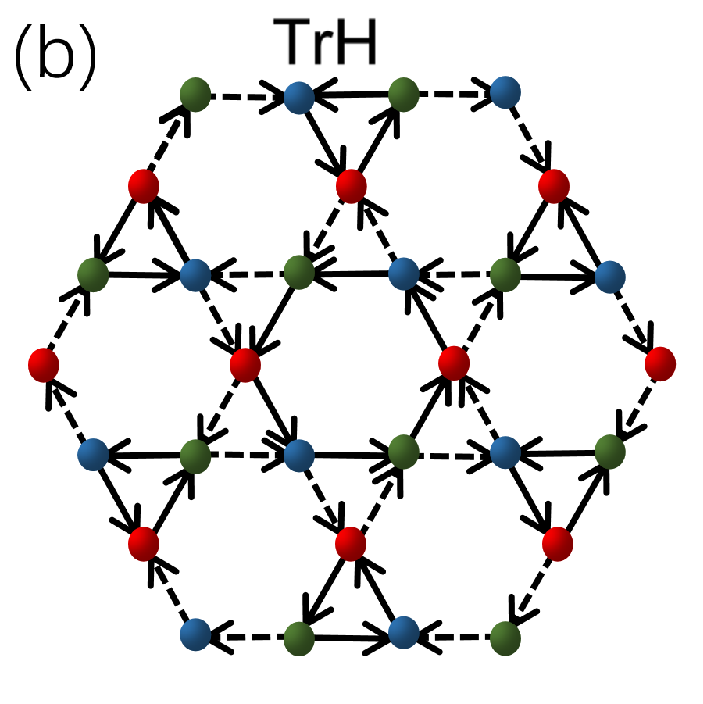} }
	\caption{\label{fig:1} CDW states in the kagome lattice showing SoD and TrH patterns.}	
\end{figure}

$H_{CDW}$ characterizes the CDW term. Two potential CDW patterns are illustrated in Figs.\ref{fig:1}(a) and \ref{fig:1}(b)~\cite{Wilson2024}. The CDW Hamiltonian encompasses both CBO and CFP terms and is given by,

\begin{align}
	H_{CDW}=&\lambda_{CBO}\sum_{\left<\textbf{ij}\right>\sigma}\eta_{\textbf{ij}}\hat{c}_{\bf i\sigma}^{\dagger}\hat{c}_{\bf j \sigma}+ \nonumber \\ &i\lambda_{CFP}\sum_{\left<\textbf{ij}\right>\sigma}\eta_{\textbf{ij}}^{*}\hat{c}_{\bf i  \sigma}^{\dagger}\hat{c}_{\bf j \sigma} +H.c., 
\end{align}
where $\eta_{\bf ij}$/$\eta_{\bf ij}^*$ are obtained from Fig.~\ref{fig:1}, with
$\eta_{\bf ij}=1$ for the solid bond and $\eta_{\bf ij}=-1$ for the dashed one. $\eta_{\textbf{ij}}^{*}$ takes the value $1$ for bonds aligned with the arrow, and $-1$ when against. The formation of SoD pattern typically implies a reduction in crystal symmetry, and the corresponding CFP phase exhibits $C_3$ rotational symmetry~\cite{Wilson2024}.

For the superconducting pairing term, $H_{SC}$, we consider isotropic $s$-wave pairing described by,
\begin{equation}
	H_{SC}=\Delta_0\sum_{{\bf k}\alpha}(\hat{c}_{\bf k \alpha \uparrow}^{\dagger}\hat{c}_{\bf -k \alpha \downarrow}^\dagger+H.c.),
\end{equation}
where $\alpha$ is the index for the sublattice.

The total Hamiltonian in the momentum space can be written as	
\begin{align}	
	H=& -t \sum_{\textbf{k}\left< \alpha  \alpha^{\prime}  \right>  \sigma} [\hat{c}_{\bf k \alpha \sigma}^{\dagger}\hat{c}_{\bf k \alpha^{\prime} \sigma}e^{-i\textbf{k}(\textbf{r}_{\alpha}-\textbf{r}_{\alpha^{\prime}})}+H.c.] \nonumber \\ &-\mu\sum_{\textbf{k}\alpha \sigma}\hat{c}_{\bf k \alpha \sigma}^{\dagger}\hat{c}_{\bf k \alpha \sigma}+\sum_{\textbf{k}\alpha }(\Delta \hat{c}_{{\bf k} \alpha \uparrow}^{\dagger}\hat{c}_{{\bf -k} \alpha \downarrow}^\dagger+H.c.) \nonumber	\\
	&+[ \lambda_{CBO}\sum_{\textbf{k} \left<\alpha  \alpha^{\prime}  \right> \sigma} \eta_{\bf \alpha \bf  \alpha^{\prime}  } \hat{c}_{\bf k \alpha \sigma}^{\dagger} \hat{c}_{\bf k  \alpha^{\prime} \sigma}e^{-i\textbf{k}(\textbf{r}_{\alpha}-\textbf{r}_{\alpha^{\prime}})} \nonumber \\
	&+i\lambda_{CFP}  \sum_{\bf k \left< \bf \alpha \bf \alpha^{\prime}  \right> \sigma} \eta_{\bf \alpha \bf \alpha^{\prime}}^{*} \hat{c}_{\bf k \alpha \sigma}^{\dagger} \hat{c}_{\bf k \alpha^{\prime} \sigma}e^{-i\textbf{k}(\textbf{r}_{\alpha}-\textbf{r}_{\alpha^{\prime}})} ]+H.c.				
\end{align}

The above Hamiltonian can be expressed as the matrix form with $H =\sum_{\bf k}\Psi _{{\bf k}}^{\dagger}\hat{M}_{\bf k}\Psi _{{\bf k}}$. $\hat{M}_{\bf k}$ is a $24\times 24$ matrix. The basis vectors $\Psi _{{\bf k}}$ is expressed as,
\begin{align}
	\Psi_{k \alpha }= (c_{\mathbf{k} 1, \uparrow}^{\dagger}, \cdots, c_{\mathbf{k} 12, \uparrow}^{\dagger}, c_{-\mathbf{k} 1, \downarrow}, \cdots, c_{-\mathbf{k} 12, \downarrow})^T. 
\end{align}

The spectral function can be obtained through diagonalizing the above Hamiltonian: 
\begin{equation}
	A(\textbf{k},\omega)=-\frac{1}{\pi} {\rm Im} \sum_{i=1}^{12}\sum_{n}[\frac{u_{in}(\textbf{k})u_{in}^{*}(\textbf{k})}{\omega- E_n(\textbf{k})+i\Gamma}+\frac{u_{jn}(\textbf{k})u_{jn}^{*}(\textbf{k})}{\omega+ E_n(\textbf{k})+i\Gamma}],
	\label{eq23}
\end{equation}
with $j=i+12$.

The density of states can be obtained by the summation of the spectral function over all momenta within the Brillouin zone as follows:
\begin{equation}
	\begin{aligned}
		\rho \left( \omega \right)&=\sum_{\bf k} A \left( {\bf k},\omega \right).
	\end{aligned}
\end{equation}

To analyze the topological features of the system, we compute the Chern number, which characterizes the system's topology. The Chern number is defined as \cite{jpsj.74.1674},
\begin{eqnarray}
	\label{EQ:chern}
	C &=& \frac{1}{2\pi i} \sum_{\bf k} \tilde{F}_{xy} \left({\bf {k}}\right).
\end{eqnarray}
where $\tilde{F}_{xy}$ represents the lattice field strength. This field strength is computed as,
\begin{eqnarray}
	\tilde{F}_{xy} \left({\bf {k}}\right) \equiv \ln \frac{ U_x\left({\bf k}\right) U_y\left({\bf k} + {\hat{x}}\right) }{U_x\left({\bf k} + {\hat{y}}\right) U_y \left({\bf k}\right)},
\end{eqnarray}
with
\begin{eqnarray}
	U_{\alpha}\left({\bf k}\right) = \frac{\det \Phi^{\dagger}\left({\bf k}\right) \Phi\left({\bf k} + {\hat{\alpha}} \right)}{\left| \det \Phi^{\dagger}\left({\bf k}\right) \Phi\left({\bf k} + {\hat{\alpha}} \right) \right|}.
\end{eqnarray}
Here, $\Phi(\mathbf{k})$ is a $24 \times 12$ matrix constructed by collating the column vectors of the occupied eigenstates, represented as $\Phi({\bf k})=(u_{1\bf{k}},u_{2\bf{k}},\cdots,u_{12\bf{k}})$. This matrix plays a crucial role in determining the topological invariants of the system by encompassing the characteristics of the occupied electronic states across the Brillouin zone.

\section{Results and Discussion}

\begin{figure}
	\centering
	\subfigure{\includegraphics[width=0.45\textwidth]{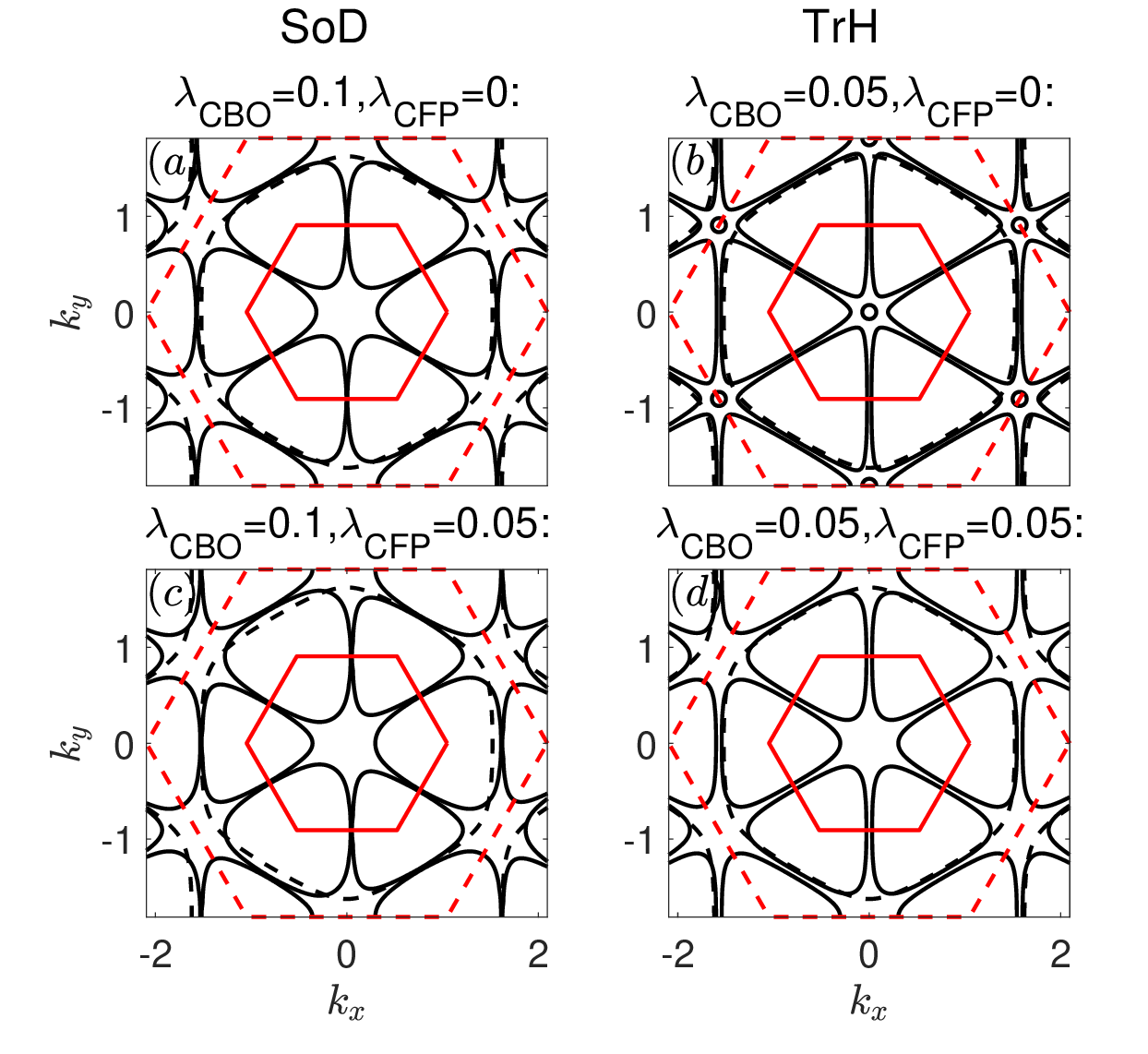} }
	\caption{\label{fig:2} Normal-state Fermi surfaces affected by various CDW orders. The dashed and solid hexagons represent the original and reduced Brillouin zones, respectively. Similarly, the dashed and solid lines depict the original and renormalized Fermi surfaces.}	
\end{figure}

We initially examined the influence of CDW ordering on the electronic structure in the normal state. In the $V$-based kagome material, the Fermi level is positioned slightly above the Van Hove singularity points, and the normal state Fermi surface forms a large pocket around the $\Gamma=(0,0)$ point~\cite{PhysRevB.106.014501,PhysRevB.109.054504}. With the introduction of CDW ordering, the Brillouin zone is reduced to one quarter of its original size. The previously observed Fermi surface folds into this reduced Brillouin zone and undergoes renormalization due to the CDW effects. Our numerical results, depicted in Fig.~\ref{fig:2}, illustrate variations in the Fermi surface across different CDW ordering schemes; the reduced Brillouin zone is outlined by a solid hexagonal boundary. Regardless of the CDW pattern examined, the initial large Fermi surface pocket transforms into six smaller pockets localized at the vertices of the reduced Brillouin zone. Despite these general trends, further distinctions exist among the results. Specifically, under the SoD pattern, the system retains a metallic state, with sizeable Fermi pockets still present. Conversely, when a CFP order is included, these pockets become distorted, with the inversion symmetry being broken. In contrast, for the TrH pattern with only CBO ordering, an additional small pocket appears around the $\Gamma$ point. As the magnitude of this CBO order increases beyond $0.054$, all Fermi pockets vanish, and the system transitions to a CDW insulator state. The introduction of additional CFP order results in the disappearance of the small pocket. Unlike the SoD pattern, the TrH configuration consistently preserves both six-fold and inversion symmetries. These characteristic features could potentially be verified through future experiments, offering a method to distinguish between the SoD and TrH patterns.

\begin{figure}
	\centering
	\subfigure{\includegraphics[width=0.45\textwidth]{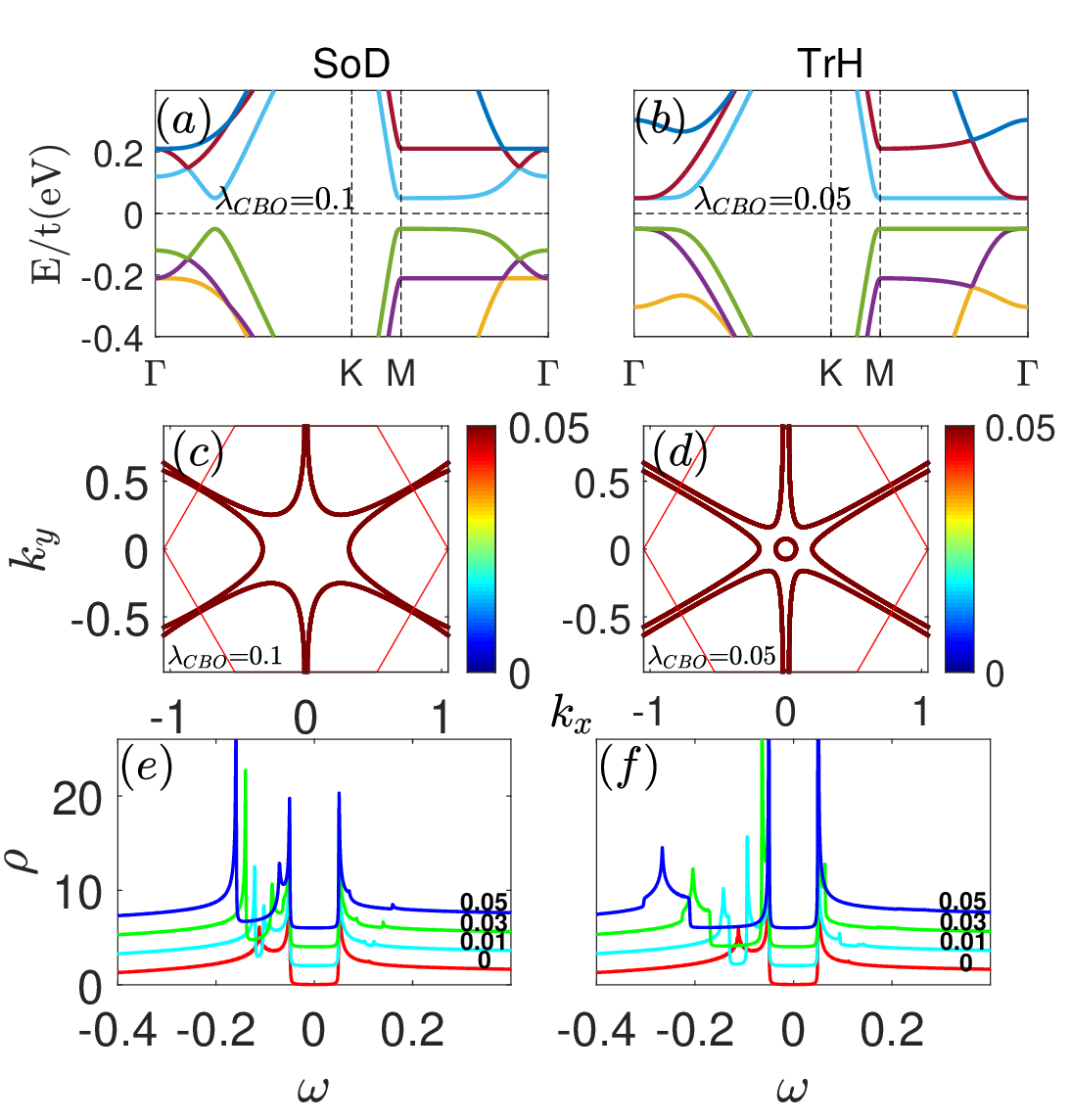} }
	\caption{\label{fig:3}  Analysis of superconducting properties in the simultaneous presence of superconductivity and CDW orders with only the CBO term. Left and right display results for SoD and TrH patterns, respectively. Panels (a) and (b) show numerical results of superconducting energy bands along highly symmetric lines in the reduced Brillouin zone; Panels (c) and (d) illustrate the effective superconducting gap along the Fermi surface; and Panels (e) and (f) present DOS spectra under varying CBO intensities.}	
\end{figure}

We subsequently analyzed the electronic structure within the region where superconducting and CDW orders coexist, focusing initially on the influence of the CBO term on the superconducting gap, set at $\Delta_0=0.05$. Energy bands along highly symmetric lines in the reduced Brillouin zone for SoD and TrH CDW states are illustrated in Figs.~\ref{fig:3}(a) and \ref{fig:3}(b), respectively. The numerical simulations reveal that the energy gaps remain consistent at $0.05$ across different Fermi momenta. Further scrutiny of the energy gaps across the renormalized Fermi surfaces for these CDW states is provided in Figs.~\ref{fig:3}(c) and \ref{fig:3}(d), confirming an isotropic superconducting gap of $0.05$ along the Fermi surface, in alignment with the preset gap value. This isotropic characteristic is maintained across varying CBO strengths, as further confirmed by the DOS spectra  with CBO strengths increasing incrementally from 0 to 0.05 (Figs.~\ref{fig:3}(e) and \ref{fig:3}(f)). These spectra demonstrate fully gapped profiles highlighted by distinct superconducting coherence peaks at the gap edges, with the positions of these peaks remaining unchanged as the CBO strengths increase.

\begin{figure}
	\centering
	\subfigure{\includegraphics[width=0.45\textwidth]{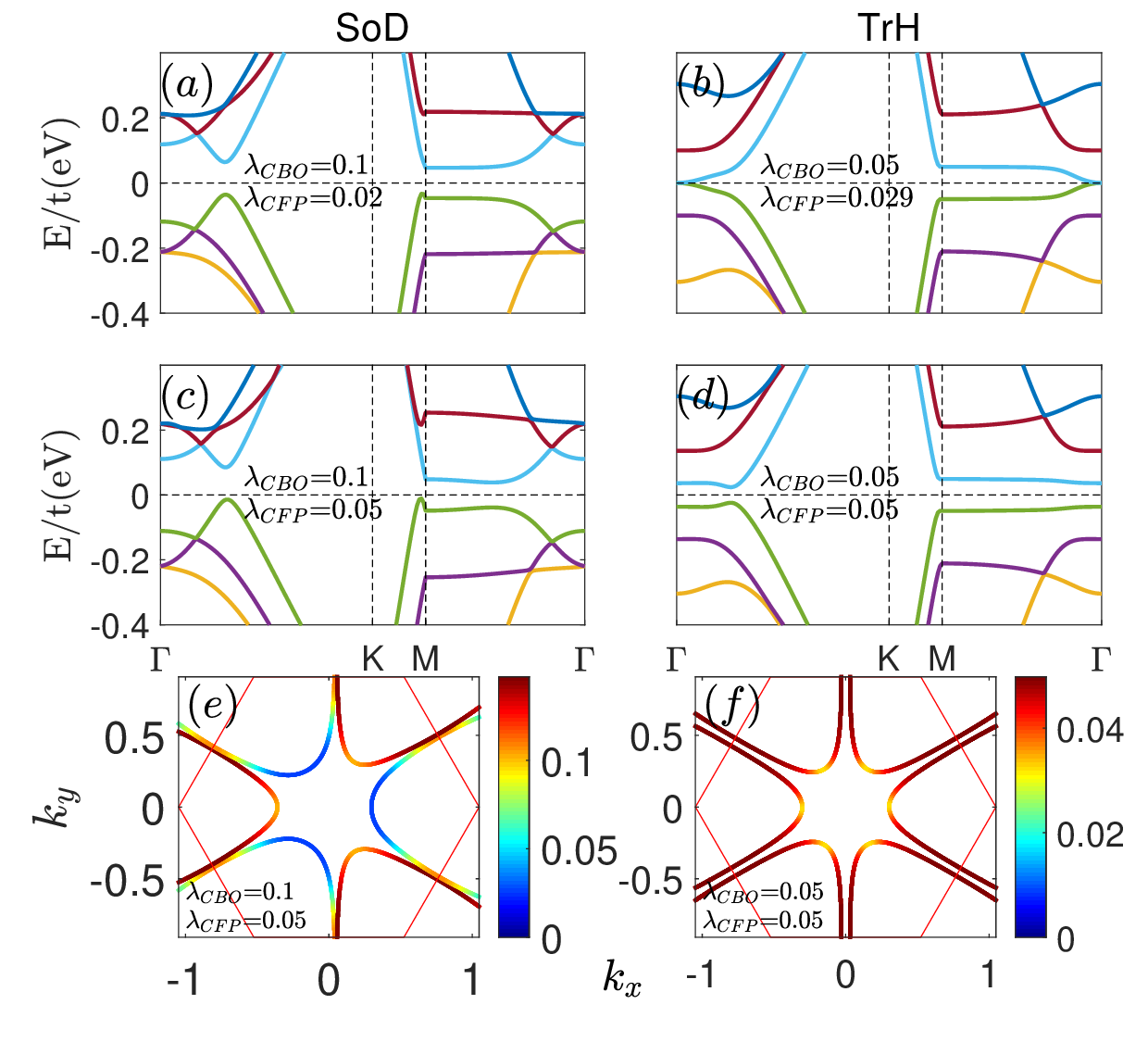} }
	\caption{\label{fig:4} Superconducting properties in the presence of both superconductivity and CDW order, incorporating CBO and CFP terms. Results for SoD and TrH patterns are shown in the left and right panels, respectively. Panels (a-d) provide the numerical results for superconducting energy bands, while Panels (e) and (f) depict the effective superconducting gap along the Fermi surface. }	
\end{figure}

Upon incorporating CFP order into the CDW scheme,  as depicted in Fig.~\ref{fig:4}, the SoD state exhibits broken time-reversal and inversion symmetries.
Although the system remains fully gapped when the CFP intensity ($\lambda_{CFP}$) is low, the energy bands begin to intersect the Fermi level as $\lambda_{CFP}$ increases. 
 For the TrH state, the energy gap closes precisely at a critical CFP intensity of $\lambda_{CFP}=0.029$. Below and above this threshold, the system retains a fully gapped state. Across all considered CFP intensities, the effective superconducting gaps along the normal state Fermi surface exhibit anisotropy in both the SoD and TrH states. These results point to intricate interactions between superconducting and CDW orders, possibly providing detailed insights into the electronic structures of these materials.

\begin{figure}
	\centering
	\subfigure{\includegraphics[width=0.45\textwidth]{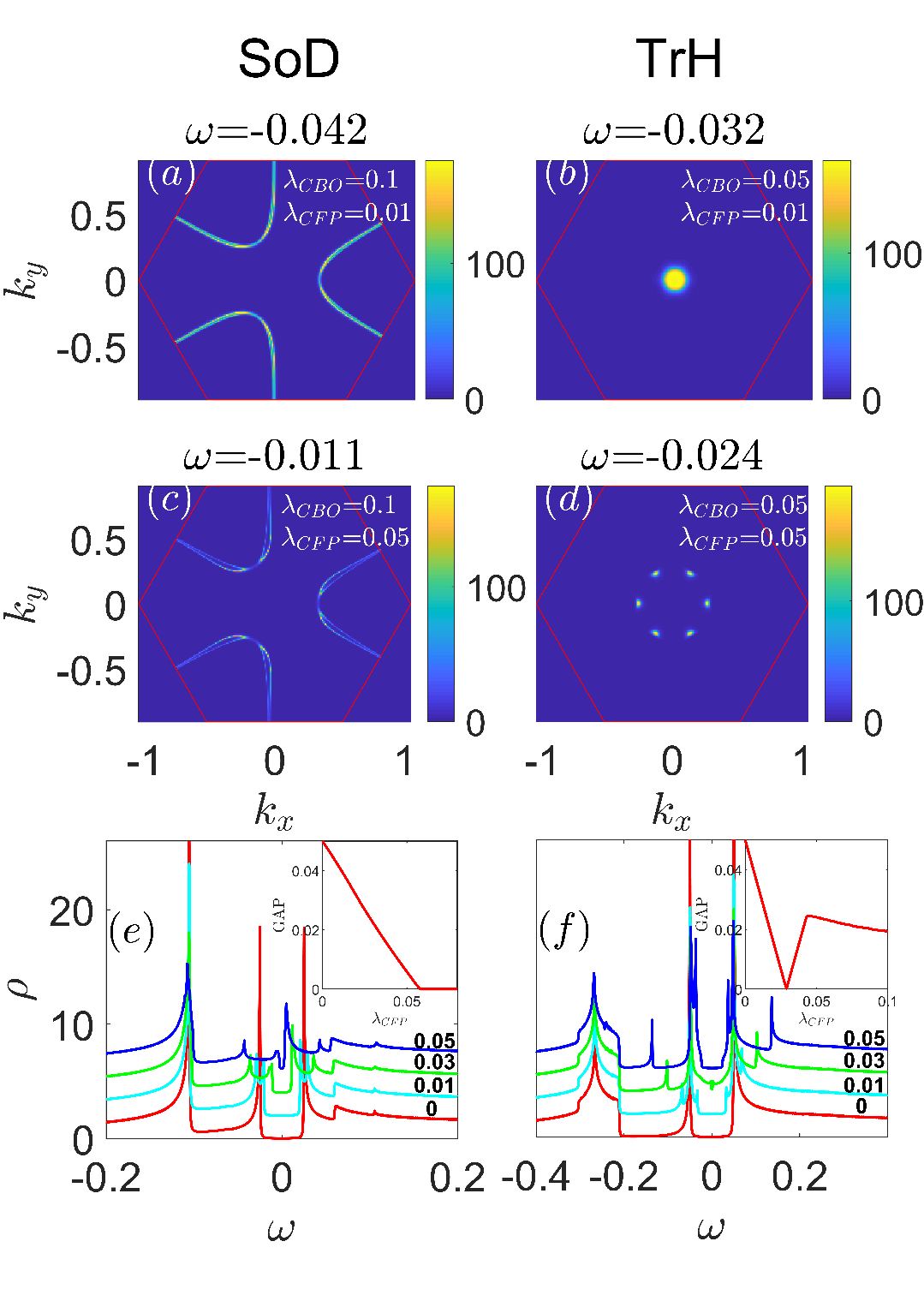} }
	\caption{\label{fig:5} Single-particle spectra in regions where superconductivity and CDW order coexist, featuring both CBO and CFP terms. Results for SoD and TrH patterns are shown in the left and right panels, respectively. Panels (a-d) present spectral function spectra with different CBO and CFP intensities. Panels (e) and (f) display DOS spectra with varying CFP intensities, and the insets of panels (e) and (f) plot the effective superconducting gaps as functions of CFP intensities. }	
\end{figure}

In experimental setups, quasiparticle energy bands can be directly examined via ARPES measurements. Theoretically, ARPES spectrum is characterized using the spectral function. The intensity plots of the spectral function, for varying CFP intensities at a constant energy, are illustrated in Figs.~\ref{fig:5}(a)-\ref{fig:5}(d). These energies are taken as the minimum quasiparticle energy at the gap edge. for all parameters considered, the spectral functions display non-uniform profiles, with notably high values at certain points, indicating minimal quasiparticle gaps at these locations.

The DOS spectra with different CFP intensities are presented in Figs.~\ref{fig:5}(e)-\ref{fig:5}(f), where the energy gap is identified by the location of the superconducting coherence peaks in these spectra. Notably, the quasiparticle gap in both CDW states is highly dependent on the CFP intensity, though the manner in which it evolves differs significantly between the two CDW patterns. In the SoD state, the energy gap decreases monotonously with increasing CFP intensity, concluding with the disappearance of coherence peaks at intensities greater than 0.052. Conversely, in the TrH state, the gap decreases until a CFP intensity of 0.029, at which point it vanishes, while it reemerges and increases as the CFP intensity continues to rise.

\begin{figure}
	\centering
	\subfigure{\includegraphics[width=0.45\textwidth]{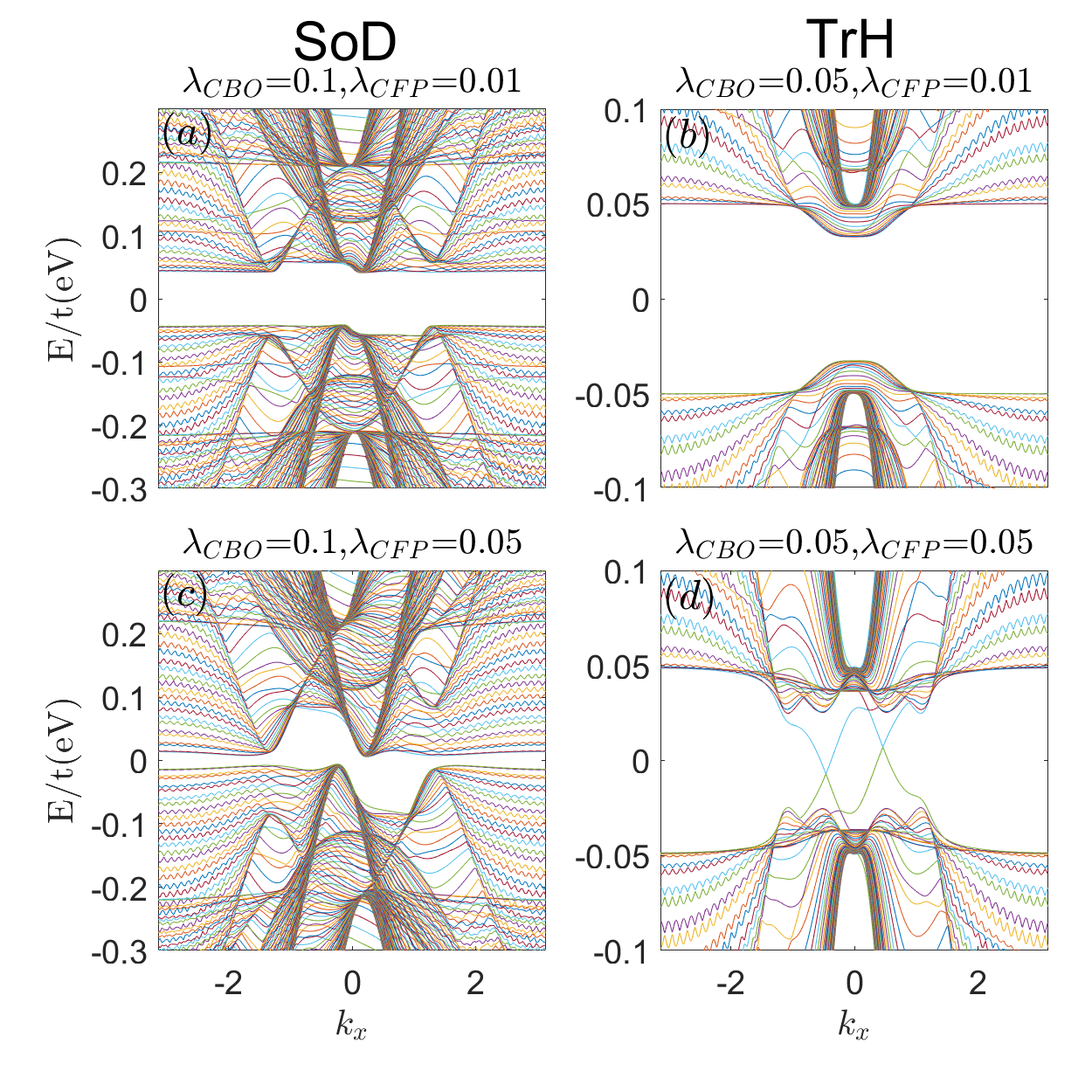} }
	\caption{\label{fig:6} Quasiparticle energy bands in the superconducting state as a function of $k_x$, considering the open boundary condition along the $y$-direction under different CFP intensities.  }	
\end{figure}

Such behavior—where the energy gap closes and subsequently reopens—suggests a potential topological phase transition within the TrH state. To probe this possibility, we calculated the Chern number for the system, finding it transitions from zero when $\lambda_{CFP} < 0.029$ to two when $\lambda_{CFP} > 0.029$. For the SoD state, the Chern number remains zero across all examined parameters. Further verification of topological characteristics was conducted through computational analysis of the system's edge states, considering a mixed open (along the y-direction) and periodic (along the x-direction) boundary condition setup, as demonstrated in Figs.~\ref{fig:6}. In the SoD state, the presence of a small CFP intensity preserves an energy gap with no edge states. However, as CFP intensity increases, the energy gap closes, aligning with the DOS observations. In contrast, for the TrH state, edge states emerge as CFP intensity rises, indicating non-trivial topological properties, consistent both with the observed edge states in the energy bands and the calculated Chern numbers.

The overall impact of the CDW term on the superconducting gap can be elucidated through an analysis of symmetry within the system. In the context of BCS-type pairing, a spin-up quasiparticle with momentum ${\bf k}$ pairs with a spin-down quasiparticle at momentum $-{\bf k}$ to form a Cooper pair. In the absence of CFP order, the electronic structures for spin-up and spin-down quasiparticles at ${\bf k}$ and $-{\bf k}$, respectively, are identical, owing to the preservation of time-reversal symmetry. This symmetry facilitates pairing across all quasiparticles near the Fermi level in the superconducting state. Conversely, the presence of CFP order breaks this symmetry, resulting in inequivalent states for opposite momenta and spins, and leads to an anisotropic superconducting gap. Specifically, in the SoD state, CFP order also disrupts inversion symmetry, causing the energy bands to lack inversion symmetry ($E_{\bf k} \neq E_{-\bf k}$). This disruption, coupled with the inherent particle-hole symmetry of the superconducting state ($E_{\bf k} = -E_{-\bf k}$), results in an asymmetric superconducting energy band configuration relative to the Fermi level, thereby enhancing gap anisotropy.

At last, we would remark the significance of our present work. Firstly, by studying the electronic structure in states where CDW and superconducting orders coexist, and considering various CDW patterns, our numerical findings aid in clarifying and identifying distinct CDW configurations. Secondly, our results confirm that the superconducting gap becomes anisotropic—and potentially nodal—upon incorporating CFP order, offering a theoretical basis to understand the varied and sometimes conflicting energy gap observations reported by different experimental teams. Thirdly, the discovery that CFP order can catalyze a transition to topological superconductivity implies that the system might host Majorana excitations, presenting a promising avenue for future research into topological superconductivity and its applications. These findings underscore the potential for novel quantum states and functionalities in kagome superconductors, meriting in-depth experimental and theoretical exploration.

\section{summary}

In summary, we investigated the electronic structure and effective superconducting gaps in kagome superconductors within regions where CDW and superconducting orders coexist. We considered two distinct CDW patterns, namely the SoD and the TrH. For both patterns, CBO order and CFP order were analyzed, corresponding to real and imaginary bond orders, respectively. Within the superconducting state, characterized by isotropic $s$-wave pairing symmetry, the effective superconducting gap remains isotropic with only CBO order present; however, it becomes anisotropic with the introduction of CFP order. Furthermore, we explored topological properties within the superconducting state and identified a topological phase transition in the TrH configuration. This transition marks the evolution from a topologically trivial to a nontrivial state, as evidenced by a change in the Chern number to $C=2$ upon increasing the CFP intensity.

\providecommand{\noopsort}[1]{}\providecommand{\singleletter}[1]{#1}%

\end{document}